\documentclass[preprint,12pt]{elsarticle}

\usepackage{epsfig}
\usepackage{amsmath,amssymb,bm}

\journal{Physics Letters B}

\begin{document}

\begin{frontmatter}
\title{The role of finite-size effects on the spectrum of equivalent photons in proton-proton collisions at the LHC.
}

\author{Mateusz Dyndal}
\address{AGH Univ. of Science and Technology, Cracow, Poland \\
CEA Saclay, Irfu/SPP, Gif-sur-Yvette, France}
\author{Laurent Schoeffel}
\address{CEA Saclay, Irfu/SPP, Gif-sur-Yvette, France}

\begin{abstract}
Photon-photon interactions represent an important class of physics processes at the LHC, where quasi-real photons
are emitted by both colliding protons. These reactions can result in 
the exclusive production of a final state $X$,  $p+p \rightarrow p+p+X$.
When computing such cross sections, it has already been shown that finite size effects of colliding protons
are important to consider for a realistic estimate of the cross sections.
These first results have been essential in understanding the physics case of heavy-ion collisions in the low invariant mass
range, where heavy ions collide to
form an exclusive final state like a $J/\Psi$ vector meson.
In this paper, our purpose is to present some 
calculations that are valid also for the exclusive production of high masses final states
in proton-proton collisions, like the
production of a pair of $W$ bosons or the Higgs boson. Therefore, we propose a complete
treatment of the finite size effects of incident protons irrespective of the mass range explored in the collision.
Our expectations are shown to be in very good agreement with existing experimental data obtained at the LHC.
\end{abstract}


\end{frontmatter}

\section{Introduction}

A significant fraction of of proton-­proton collisions at large
energies involves quasi-­real photon interactions.
This fraction is dominated by elastic scattering, with a single Born-level photon exchange.
The photons can also be emitted by both protons, where a variety of
central final states can be produced.
The proton-proton collision is then transformed into a photon-photon interaction
and the protons are deflected at small angles. At the LHC, these reactions
 can be measured at the
energies well beyond the electroweak energy scale. This offers an interesting field of research linked to photon-photon interactions,
 where the available effective luminosity is small, relative to parton-parton interactions, but is compensated
by better known initial conditions and usually simpler final states. Indeed,
for high energetic proton-proton collisions, at a center of mass energy $s$, the idea is  to search for
the exclusive production of a final state $X$ through the reaction  $p+p \rightarrow p+p+X$. Therefore, the initial state 
formed by both photons is well-defined, while the final state formed by $X$ with no other hadronic activity is much simpler than
in a standard inelastic proton-proton  interaction. In the following, we write this reaction as 
$p p (\gamma \gamma) \rightarrow p p X$.

In order to compute the cross section for the process $p p (\gamma \gamma) \rightarrow p p X$,
we need to consider that each of the two incoming protons emits a quasi-real photon which fuse to give a centrally produced final state $X$
($\gamma+\gamma \rightarrow X$).
This calculation relies on the so-called
 equivalent photon approximation (EPA) \cite{fermi,ww,t1,t2,Budnev:1973tz}.
The EPA is based on the property that
the electromagnetic (EM) field of a  charged particle, here a proton, moving at high velocities becomes more and more transverse
with respect to the direction of propagation.  As a consequence, an observer in the laboratory frame can not distinguish 
between the EM field of the relativistic proton and its transverse component,
which can be labeled as the transverse EM field of equivalent photons.
This implies that the total cross section of the reaction $p p (\gamma \gamma) \rightarrow p p X$ can be approximately described as a photon-photon fusion cross section ($\gamma \gamma \rightarrow X$)
folded with the equivalent photon distributions $f(.)$ for the two protons
\begin{equation}
\sigma(p+p \rightarrow p+p+ X) =\int \int f(\omega_1)f(\omega_2) \sigma_{\gamma \gamma \rightarrow X}(\omega_1,\omega_2) 
\frac{d\omega_1}{\omega_1} \frac{d\omega_2}{\omega_2},
\label{start}
\end{equation}
where $\omega_{1,2}$ represent the energies of the photons and are integrated over.
For each photon, the maximum energy is obviously the energy of the incident proton $\sqrt{s}/2$. However, there is also the 
constraint that the highest available energy for one photon is of the order of the inverse Lorentz contracted radius of the proton, 
$\gamma/r_p$,
where $r_p$ represents the proton radius.
Let us note that the two photon center-of-mass energy squared is $W_{\gamma \gamma}^2 = 4\omega_1\omega_2$, and the rapidity of the two photons system is defined as
$y_{\gamma \gamma} = 0.5 \ln[{\omega_1}/{\omega_2}]$.

In equation (\ref{start}), the photon distributions $f(.)$ are already integrated over the virtuality  ($Q_{1,2}^2$)
of the photons. As this dependence is of the order of $1/Q_{1,2}^2$, this justifies the approximation that both
photons are quasi-real.

We can remark that for practical issues, the situation may be more complex. Indeed, each proton can either survive and, then, is scattered at  a small angle, as considered above. This is the case of elastic emission. 
Elastic two-photon processes yield very clean event topologies at the LHC: two very forward protons measured away from the interaction point
and a few centrally produced particles (forming the final state $X$).
But, it is also possible that one or both protons  dissociate into a hadronic state. This is the case of inelastic emission. In this paper, we restrict the discussion to the elastic case.

Let us note also that the calculations presented in this paper are commonly used for heavy-ion collisions, where the EPA approximation 
can be applied similarly. Only the charges and the radii of the incident particles are modified in this case.

\begin{figure}[hbtp]
\centering
  \includegraphics[scale=0.4]{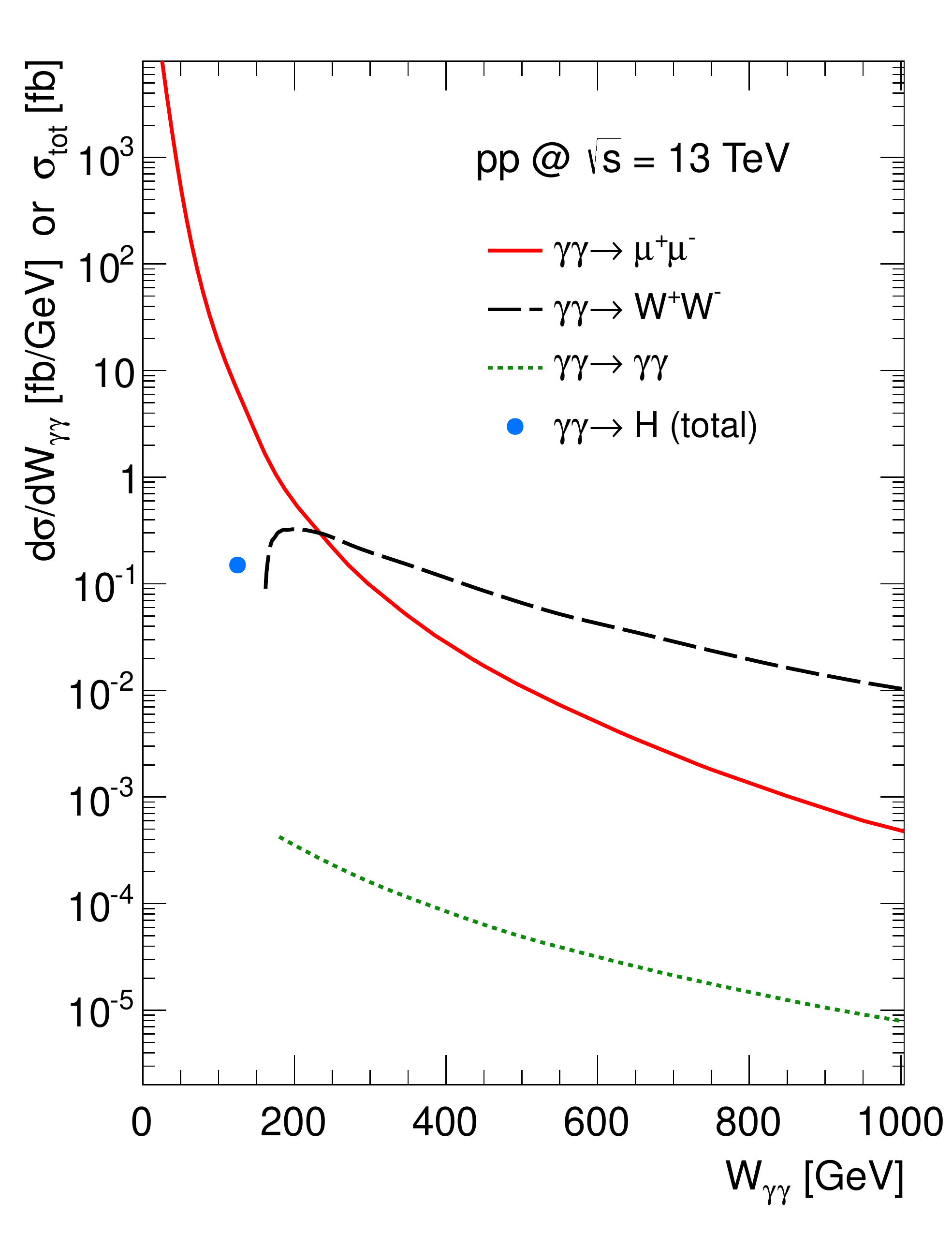}
   \caption[]{Cross sections of various  processes $p p (\gamma \gamma)\rightarrow p p X$, differential 
in the photon-photon center of mass energy. For the exclusive Higgs production, the total cross section is shown.
The exclusive production of pairs of photons has been
generated  at large $W_{\gamma \gamma}$ where the cross section
is dominated by one-loop diagrams involving $W$ bosons.
}   
  \label{figxs}
\end{figure}

Previous studies have been done using equation (\ref{start}) in order to compute cross sections at LHC energies for various 
 photon-photon processes in proton-proton collisions,
$p p (\gamma \gamma)\rightarrow p p X$, corresponding to different final states $X$ \cite{d'Enterria:2008sh,deFavereaudeJeneret:2009db}.
Some results are displayed in figure
\ref{figxs}. 
The exclusive production of pairs of muons and pairs of $W$ bosons have been
generated using the {\sc Herwig++} generator \cite{Bahr:2008pv}. The exclusive production of pairs of photons has been
generated using the {\sc FPMC} generator \cite{Boonekamp:2011ky} at large $W_{\gamma \gamma}$ where the $\gamma \gamma \rightarrow 
\gamma \gamma$ cross section
is dominated by one-loop diagrams involving $W$ bosons \cite{inan}. Finally, the 
exclusive production of the Higgs boson is computed according to higgs effective field theory ({\sc HEFT}) \cite{eft}.
Obviously, this last reaction appears as a point in figure \ref{figxs}, representing the total cross section, at the Higgs mass. 

In this paper, our purpose is to generalize equation (\ref{start}) to the physics case where the impact parameter
dependence of the interaction can not be neglected \cite{Szczurek:2014yba}. In particular, we show that this approach is needed when we take in
consideration
 the finite size of colliding protons (or  heavy-ions) in the calculations.
This is not new in the sense that these finite size effects have 
already been encoded in the {\sc Starlight} Monte-Carlo \cite{Klein:2003vd} dedicated to heavy-ion collisions.
Let us note that {\sc Starlight} is not restricted to photon-photon interactions but can also be used in photon-Pomeron configurations,
as it is done at LHCb \cite{lhcb}.
However, {\sc Starlight} is focused mainly on the low invariant mass domain around the mass of the $J/\Psi$,
which justifies some approximations made for example by neglecting the magnetic form factors.

In the following we develop some calculations that are valid also for the exclusive production of high masses final states
in proton-proton collisions, like the
production of a pair of $W$ bosons or the Higgs boson. Therefore, our purpose in this paper is to propose a complete
treatment of the finite size effects of incident protons irrespective of the mass range explored in the collision.
In section 2, these calculations are presented extensively. Then, results are discussed in section 3 and compared to existing measurements.

\section{Impact parameter dependent equivalent photon method}

Deriving the expression of the equivalent photon distribution of the fast moving proton without neglecting the 
impact parameter dependence  means that we determine this distribution
as a function of the energy of the photon and the  
distance $\vec{b}$ to the proton trajectory. 
This distance is defined in the plane transverse to the proton trajectory. Therefore we speak of transverse distance.
This last dependence is not present in the approach based on formula (\ref{start}).
Following calculations presented in
\cite{greiner,baltz}, the general equivalent photon distribution read
\begin{equation}
n({b},\omega) = \frac{\alpha_{EM}}{\pi^2 \omega}
\left |
\int d{k}_\perp {k}_\perp^2
 \frac{F(k_\perp^2+\frac{\omega^2}{\gamma^2})}{k_\perp^2+\frac{\omega^2}{\gamma^2}}
J_1(b k_\perp)
\right |^2 
\label{photond1}
\end{equation}
where $\gamma$ is the Lorentz contraction factor, 
 $\omega$ and $\vec{k}_\perp$ represent the energy and transverse momentum of photons respectively.
In this expression, $F(.)$ is the proton form factor, electric and magnetic, that we develop explicitly below.
Let us note that $n({b},\omega)$ depends only on the modulus of the impact parameter as
obviously this quantity does not depend on the orientation of $\vec{b}$.
We can introduce the virtuality of the photon $Q^2=-k^2=k_\perp^2+\frac{\omega^2}{\gamma^2}$. Then,
expression (\ref{photond1}) becomes
\begin{equation}
n({b},\omega) = 
\frac{\alpha_{EM}}{\pi^2 \omega}
\left |
\int d{k}_\perp {k}_\perp^2
 \frac{F(Q^2)}{Q^2}
J_1(b k_\perp)
\right |^2,
\end{equation}
After developing the complete expression of the form factor $F(.)$, we get
\begin{equation}
n({b},\omega)= \frac{\alpha_{EM}}{\pi^2 \omega}
\left |
\int d{k}_\perp {k}_\perp^2
\frac{G_E(Q^2)}{Q^2} \left [   
(1-x) \frac{4m_p^2+Q^2\mu_p^2}{4m_p^2+Q^2}+\frac{1}{2} x^2 \frac{Q^2}{{k}_\perp^2}\mu_p^2
\right ]^{\frac{1}{2}}
J_1(b k_\perp)
\right |^2,
\label{toto1}
\end{equation}
where $x$ is the energy fraction of the proton carried by the photon, given by $x=2\omega/\sqrt{s}$.
Let us note that the electromagnetic coupling strength $\alpha_{EM}$ is taken to be $\alpha_{EM}(Q^2 \simeq 0$ GeV$^2)=1/137.036$ 
throughout our calculations,
following the property that the photons entering the interaction are quasi-real (see section 1).

\begin{figure}[!ht]
\centering
  \includegraphics[scale=0.4]{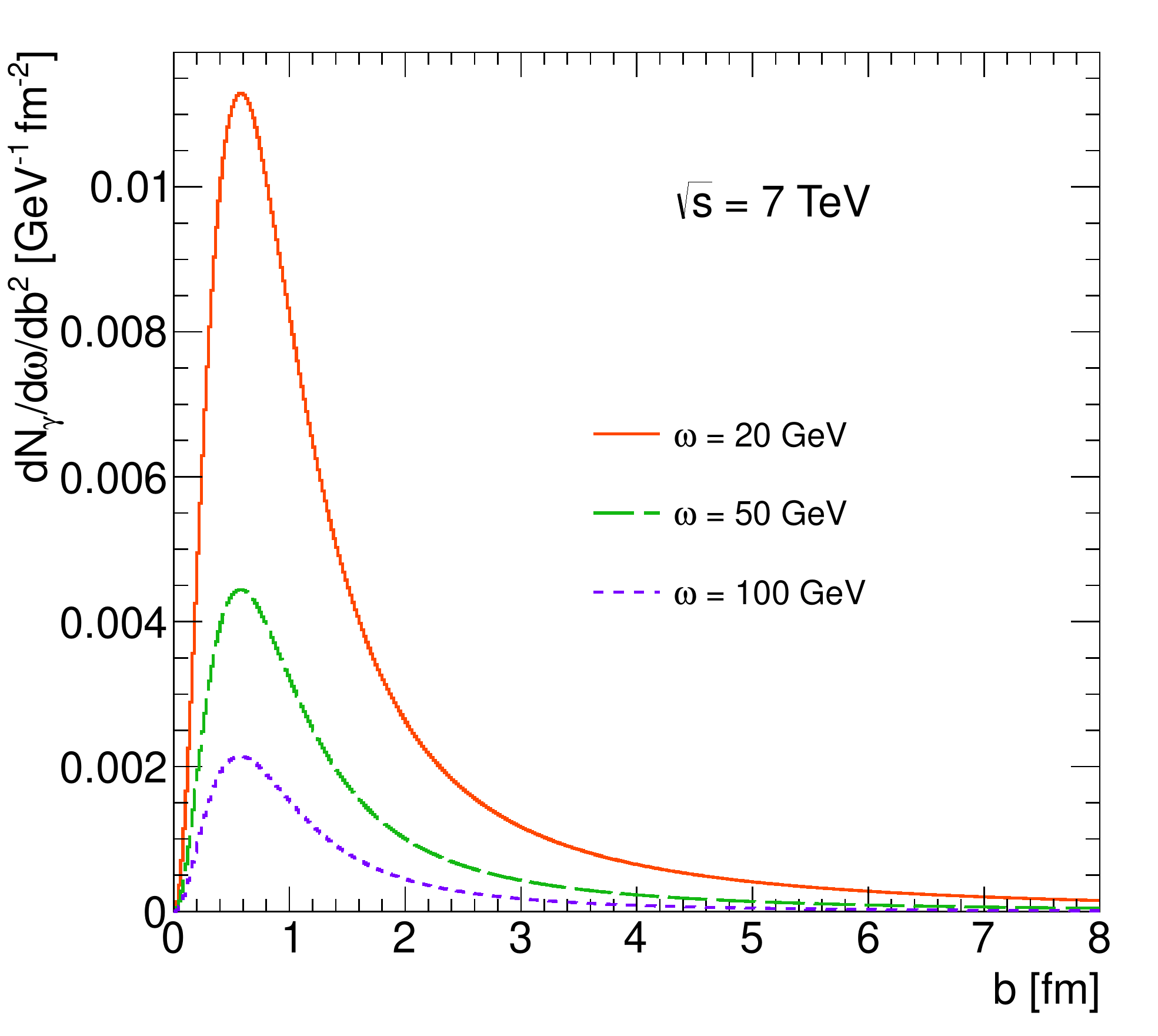}
   \caption[]{Equivalent photon distributions
of the fast moving proton
for different energies of the photon, as function of the transverse distance $b$ (see text).
}
  \label{fign}
\end{figure}

\begin{figure}[hbtp]
\centering
  \includegraphics[scale=0.5]{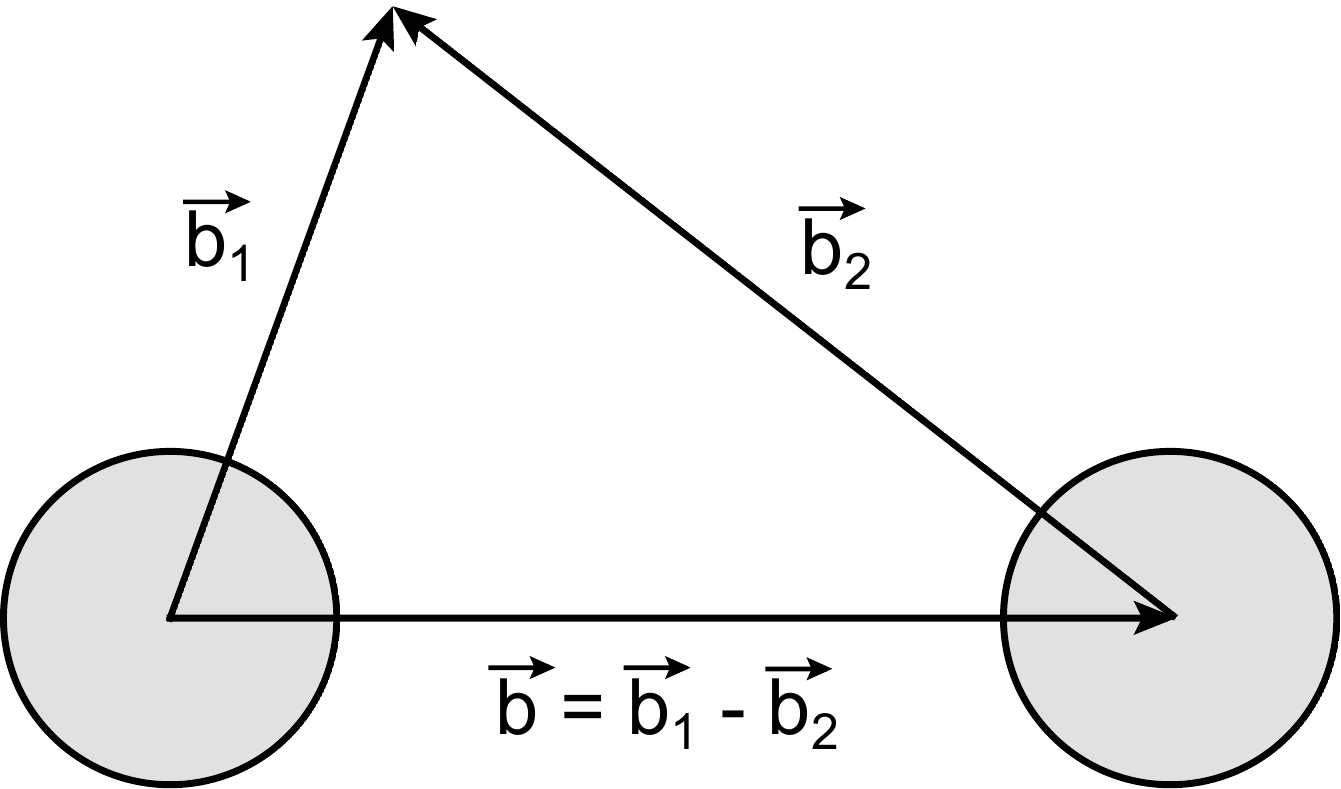}
   \caption[]{Schematic view of the two protons and the transverse distances $\vec{b}_{1}$ and  $\vec{b}_{2}$.
The difference $\vec{b}=\vec{b}_{1}-\vec{b}_{2}$  is also pictured. This is clear from this view that
the geometrical non-overlapping  condition of the two protons corresponds to $|\vec{b}_{1}-\vec{b}_{2}| > 2r_p$.}   
  \label{pict}
\end{figure}

The relation (\ref{toto1}) for $n({b},\omega)$ corresponds to the equivalent photon 
distribution (for one proton) when  the impact parameter dependence is taken into account.
Equivalent photon distributions are presented in figure \ref{fign}, as a function of the impact parameter
for different energies of the photon. 
The overall shapes of these distributions can be understood easily. At very large $b$ values, $n(b,\omega)$ 
behaves asymptotically as $\frac{1}{b}e^{-2\omega b/\gamma}$ for what concerns its $b$ dependence. At very small $b$ values,
the photon distributions are damped due to the effects of form factors and finite size of the proton.
We can remark that
equation (\ref{start}) can  be re-derived from  expression (\ref{toto1}) after replacing
$f(\omega_1)$ by the integral of
$
n(\vec{b}_1,\omega_1) 
$ for all $\vec{b}_1$,
and similarly for the second photon variables independently. Indeed
$$
f(\omega)=
\frac{e^2}{\pi \omega}
\int \frac{d^2 \vec{k}_\perp}{(2 \pi)^2} 
  \left(\frac{F(k_\perp^2+\frac{\omega^2}{\gamma^2})}{k_\perp^2+\frac{\omega^2}{\gamma^2}}\right)^2
|\vec{k}_\perp|^2,
$$
where we have used the generic expression for the form factor of the proton,
as in equation (\ref{photond1}).

The full expression (\ref{toto1}) is necessary  when we want to take into account  effects that depend directly on the 
transverse space variables of the reaction. Therefore, when we consider the finite sizes of colliding protons,
we need to do the replacement
\begin{equation}
f(\omega_1)f(\omega_2) \rightarrow \int \int n(\vec{b}_{1},\omega_1) n(\vec{b}_{2},\omega_2) 
d^2 \vec{b}_{1} d^2 \vec{b}_{2},
\label{rep}
\end{equation}
where the bounds of integrations on the transverse distances $\vec{b}_{1}$ and  $\vec{b}_{2}$ 
prevent from performing the integrations independently.
Indeed, there are important geometrical constraints to  encode:
the two photons  need to interact at the same point outside the two protons, of radii $r_p$, while the proton-halos do not overlap.
This implies minimally that $b_1 > r_p$, $b_2 > r_p$ and $|\vec{b}_{1}-\vec{b}_{2}| > 2r_p$ 
(see figure \ref{pict}).
The last condition clearly breaks the factorization in the variables $\vec{b}_{1}$ and $\vec{b}_{2}$ of
the integral  (\ref{rep}).
In these conditions, the proton radius $r_p$ is the two-dimensional radius, determined in the transverse plane, that will be taken to be $0.64\pm 0.02$, as measured in the H1 experiment \cite{Aaron:2009ac}.
Let us note that it would be possible to keep the factorization by imposing stronger constraints,
like $b_{1,2} > 2r_p$. However, this last condition prevents configurations where the two protons are very close and produce very energetic photon-photon collisions. This is not what we want.

\begin{figure}[hbtp]
\centering
  \includegraphics[scale=0.4]{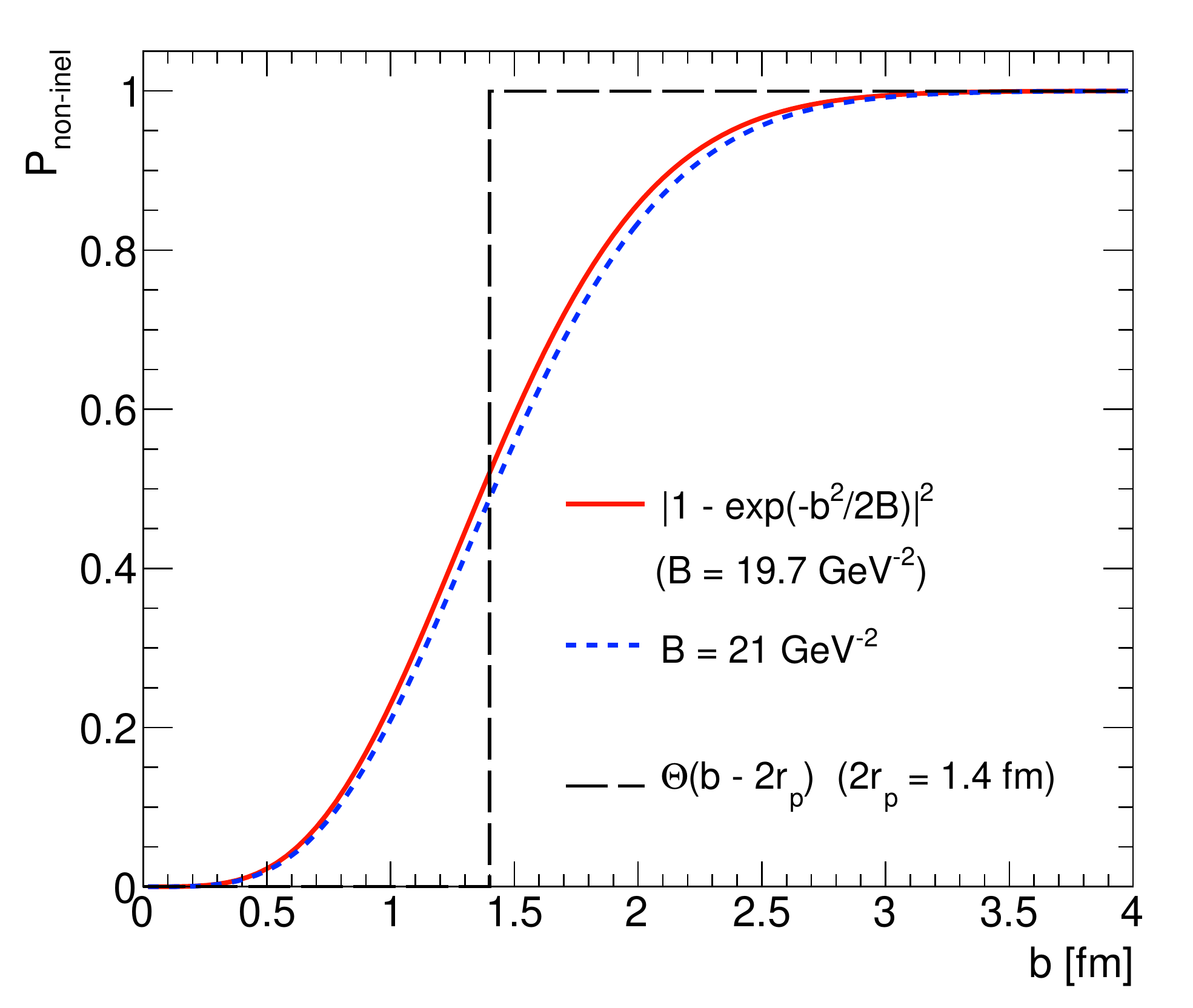}
   \caption[]{Function $P_{non-inel}(b)=\left|  1-\Gamma(b) \right|^2$ compared with 
the step function $\Theta(b-2R)$. $P(b)$ represents the probability for no inelastic interaction
in a proton-proton collision at impact parameter $b$.}   
  \label{fig3}
\end{figure}

Equation (\ref{rep}) is a first step towards encoding finite size effects. It can be refined by including
the proton-proton interaction probability, which depends explicitly on the transverse variables, 
$P_{non-inel}(|\vec{b}_{1}-\vec{b}_{2}|)$.
Then, equation (\ref{rep}) becomes
\begin{equation}
f(\omega_1)f(\omega_2) \rightarrow \int \int n(\vec{b}_{1},\omega_1) n(\vec{b}_{2},\omega_2) 
P_{non-inel}(|\vec{b}_{1}-\vec{b}_{2}|)
d^2 \vec{b}_{1} d^2 \vec{b}_{2},
\label{rep2}
\end{equation}
where the bounds of integrations are still $b_1 > r_p$, $b_2 > r_p$. The non-overlapping condition
$|\vec{b}_{1}-\vec{b}_{2}| > 2r_p$ is not needed any longer. It follows as a consequence of the 
effect of the function $P_{non-inel}(|\vec{b}_{1}-\vec{b}_{2}|)$. Indeed, this function represents the probability
that there is no interaction (no overlap) between the two colliding protons in impact
parameter space. Following \cite{Frankfurt:2006jp}, we make the natural assumption that 
a probabilistic approximation gives a reasonable estimate of the absorption effects. 
Then, we can write \cite{Frankfurt:2006jp}
$$
P_{non-inel}(b) = |1-\exp(-b^2/(2B))|^2,
$$
where the value of $B=19.7$ GeV$^{-2}$ is taken from a measurement at $\sqrt{s}=7$ TeV by the ATLAS experiment  \cite{Aad:2014dca}
(see figure \ref{fig3}).
At $\sqrt{s}=13$ TeV, we will use the extrapolated value $B=21$ GeV$^{-2}$.
In figure \ref{fig3}, we compare 
 $P_{non-inel}(b)$ with the step function $\Theta(b-2r_p)$, which is the first approximation that we have described above to quantify a non-overlapping condition
between both protons. We see that both functions are roughly comparable. However, we can expect some deviations when performing
more accurate computations of cross sections using $P_{non-inel}(b)$ in equation (\ref{rep2}), and then in equation (\ref{start}).

\section{Results}

Following the previous section, the first important issue is 
to quantify the size of the correction when we take into account the finite size of colliding protons.
We define the survival factor as
\begin{equation}
S_{\gamma \gamma}^2=\frac{
\int_{b_1>r_p} \int_{b_2>r_p} n(\vec{b}_{1},\omega_1) n(\vec{b}_{2},\omega_2) 
P_{non-inel}(|\vec{b}_{1}-\vec{b}_{2}|)
d^2 \vec{b}_{1} d^2 \vec{b}_{2}
}
{
\int_{b_1>0} \int_{b_2>0}  n(\vec{b}_{1},\omega_1) n(\vec{b}_{2},\omega_2) 
d^2 \vec{b}_{1} d^2 \vec{b}_{2}
},
\label{survival}
\end{equation}
where the numerator contains the finite size effects encoded in the function $P_{non-inel}(b)$ and dedicated bounds of the
integrations over $\vec{b}_{1}$ and $\vec{b}_{2}$, whereas the denominator represents the integral over all impact parameters 
with no constraint.

\begin{figure}[btp]
\centering
  \includegraphics[scale=0.4]{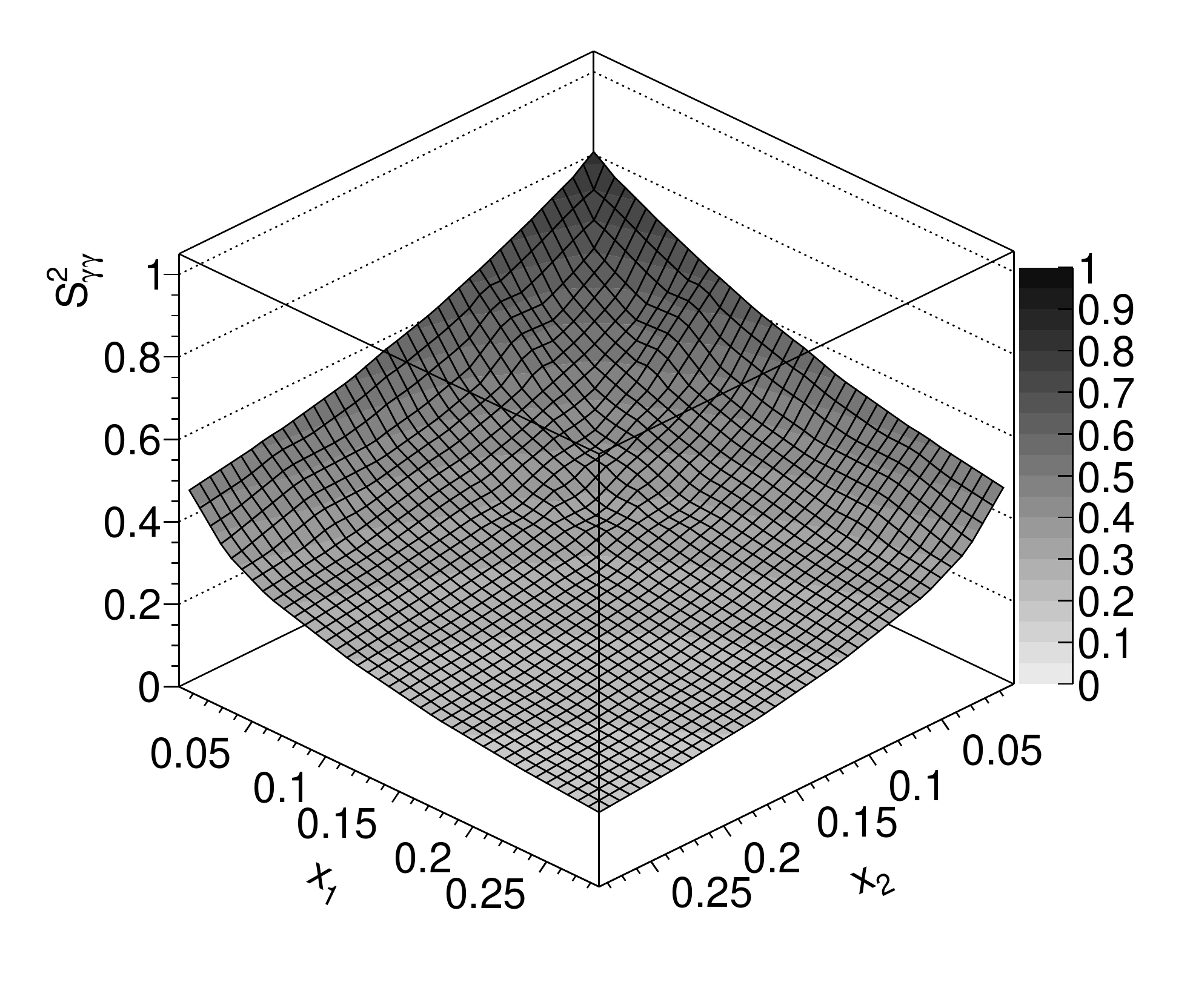}
   \caption[]{The survival factor as a function of the energy fractions of the protons carried by the interacting photons, $x_1$
and $x_2$.}   
  \label{figsxx}
\end{figure}

\begin{figure}[btp]
\centering
  \includegraphics[scale=0.4]{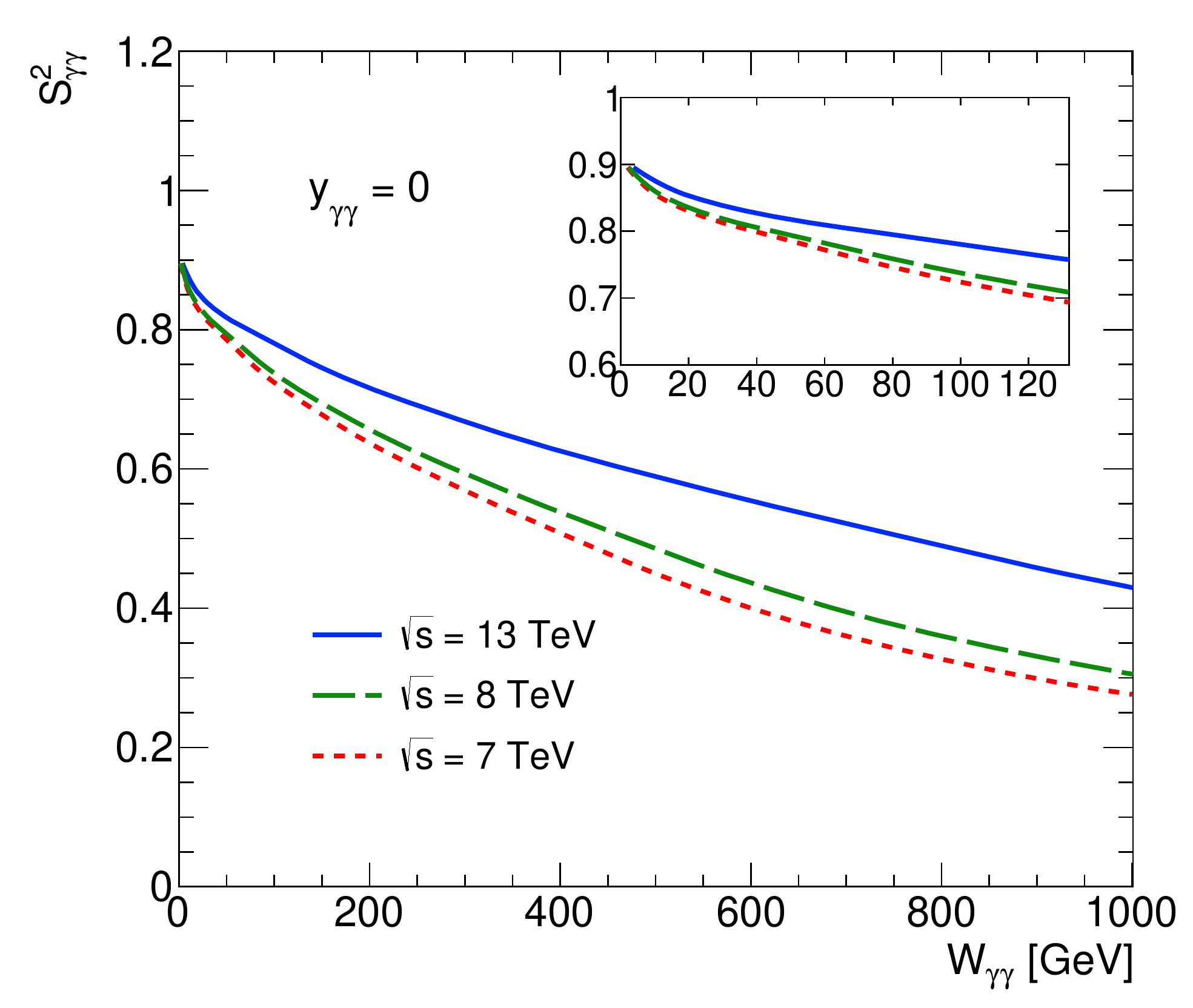}
   \caption[]{The survival factor at zero rapidity as a function of the photon-photon center of mass energy.}   
  \label{fig4}
\end{figure}

\begin{figure}[btp]
\centering
  \includegraphics[scale=0.4]{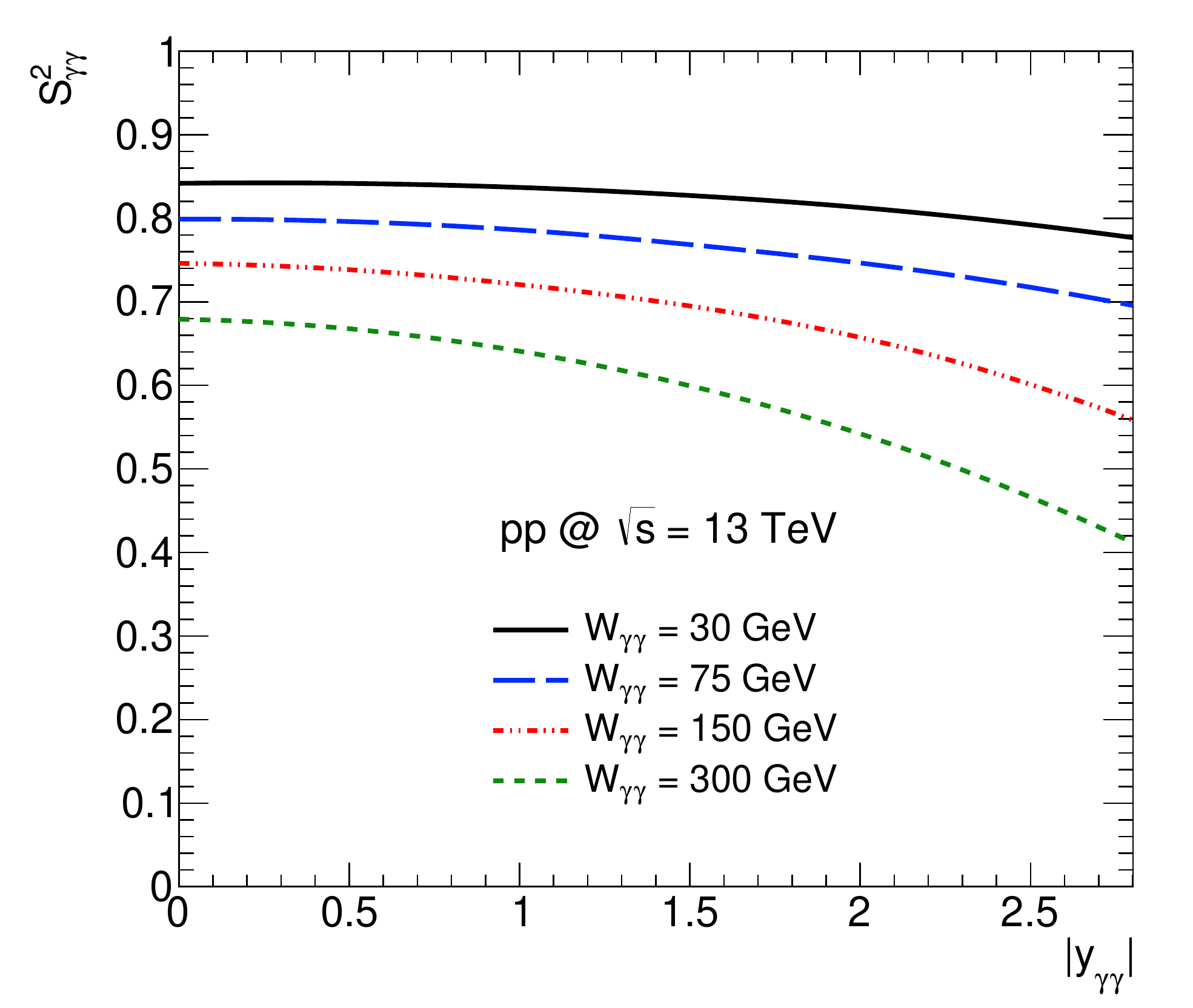}
   \caption[]{The survival factor for different the photon-photon center of mass energies displayed as
a function of the rapidity of the photon-photon system.}   
  \label{fig5}
\end{figure}

Trivially, this factor will always be smaller than unity. Then, the deviation with respect to unity will quantify the overestimation 
done when the finite size effects are neglected.
This is first illustrated in figure \ref{figsxx}, where we present the two-dimensional
dependence of $S^2_{\gamma \gamma}$ as a function of $x_1$ and $x_2$, the 
energy fractions of the protons carried by the interacting photons.
Then,
the survival factor is displayed as a function of experimentally measurable variables in figures \ref{fig4} and \ref{fig5}.
Figure \ref{fig4}  presents the behavior of the survival factor as a function of the center of mass energy of the photon-photon
system ($W_{\gamma \gamma}$) at zero rapidity.
Different curves are displayed corresponding to the different center of mass energies for the proton-proton collision.
We observe a common feature. For all curves, the survival factor is decreasing as a function of $W_{\gamma \gamma}$, to reach values of
$0.3$ at $W_{\gamma \gamma}=1$ TeV for $\sqrt{s}=7$ or $8$ TeV and $0.43$ 
at $W_{\gamma \gamma}=1$ TeV for $\sqrt{s}=13$ TeV. This is a large effect, 
due to the fact that for larger values of $W_{\gamma \gamma}$, smaller values of
$b=|\vec{b}_{1}-\vec{b}_{2}|$ are probed, and thus the integral at the numerator of  the survival factor (\ref{survival}) becomes 
smaller. Indeed, when the photon-photon energy becomes larger and larger, this is understandable that the probability of no inelastic interaction becomes smaller and smaller.
Figure \ref{fig4}  illustrates the behavior of the survival factor as a function of the rapidity of the photon-photon system, for different $W_{\gamma \gamma}$. Obviously, we observe the same effect as in figure (\ref{fig4}), that when $W_{\gamma \gamma}$ increases the survival factor decreases. In addition, this figure shows the small dependence as a function of the rapidity $y_{\gamma \gamma}$. 
Let us note that for possible
measurements at the LHC, the rapidity domain covered is close to zero. Therefore, the dependence in $y_{\gamma \gamma}$ is a marginal
effect.

As a second step, we can compute cross sections for various processes $p p (\gamma \gamma) \rightarrow p p X$
taking correctly into account the finite size effects of incident protons
and thus the survival factor. As discussed in the previous section, this requires equation (\ref{start})
with the replacement (\ref{rep2}). A set of predictions is presented 
in table \ref{tab:xsComp}, where  total cross sections are shown,
cumulative in $W_{\gamma\gamma}$ above the bounds indicated in the table. For the exclusive production of pairs of $W$ bosons, 
this is the natural bound which applies of $2M_W$.

\renewcommand{\arraystretch}{1.3}
\begin{table}[t!]
  \begin{center}
    \begin{tabular}{lccc}
      \hline \hline
      Process &\multicolumn{1}{c}{$\sigma_{tot}$} &\multicolumn{1}{c}{$\sigma_{tot} \otimes S^2_{\gamma\gamma}$} & \multicolumn{1}{c}{$<S^2_{\gamma\gamma}>$}\\ 
	\hline
    $\gamma\gamma \rightarrow H$ ($M_H=125$~GeV) & 0.15 fb & 0.11 fb & 0.74 \\
	$\gamma\gamma \rightarrow \mu^+\mu^-$ ($W_{\gamma\gamma}>40$~GeV) & 12 pb & 10 pb & 0.8 \\	
	$\gamma\gamma \rightarrow \mu^+\mu^-$ ($W_{\gamma\gamma}>160$~GeV) & 36 fb & 25 fb & 0.7 \\
	$\gamma\gamma \rightarrow W^+W^-$ & 82 fb & 53 fb & 0.65 \\
	$\gamma\gamma \rightarrow \gamma\gamma$ ($W_{\gamma\gamma}>200$~GeV) & 0.06 fb & 0.04 fb & 0.64 \\
	     \hline \hline
    \end{tabular}
  \caption{Comparison of total cross sections at $\sqrt{s}=13$ TeV for different procsesses $p p (\gamma \gamma) \rightarrow p p X$ with and without proton survival factor applied.}
  \label{tab:xsComp}
  \end{center}
\end{table}
\renewcommand{\arraystretch}{1}

Finally, we can compare our results with the experimental measurements available. Recently,
the 
CMS experiment has measured exclusive pair of muons production \cite{Chatrchyan:2011ci} and has 
reported the value of $S^2_{\gamma\gamma}$~to be $0.83\pm 0.15$ for invariant masses of the photon-photon system above $11.5$ GeV.
This is consistent with our expectations from figure (\ref{fig4}), which, convoluted with the elementary cross section in this kinematic 
range, gives a survival factor of $0.84$. In addition, 
in the analysis of the exclusive production of pairs of $W$ bosons  by the CMS experiment \cite{Chatrchyan:2013foa}, using exclusive muons production 
as a benchmark,
the measured survival factor $S^2_{\gamma\gamma}$ is found to be about $10$\% smaller that the one above
for invariant masses above $40$~GeV. This is also consistent with our expectations ($S^2_{\gamma\gamma} = 0.76$ in this kinematic domain).

\section{Conclusion}

The exclusive production of a final state $X$,  $p p (\gamma \gamma) \rightarrow p p X$,  represents an essential
class of reactions at the LHC, mediated through photon-photon interactions.
The interest of such processes is due to their well-known initial conditions and simple final state.
In this paper, we have presented a complete treatment of finite size effects of colliding protons,
needed to compute the corresponding cross sections for these reactions.
We have derived a survival factor that quantifies the deviation of the complete treatment with respect to 
no size effect.

We have shown that
the survival factor is decreasing as a function of mass of the photon-photon system ($W_{\gamma \gamma}$), to reach values of
$0.3$ at $W_{\gamma \gamma}=1$ TeV for $\sqrt{s}=7$ or $8$ TeV and $0.43$ at
$W_{\gamma \gamma}=1$ TeV for $\sqrt{s}=13$ TeV. This is a large effect, 
due to the fact that for larger values of $W_{\gamma \gamma}$, the probability of no inelastic interaction becomes smaller and smaller
and so the survival factor.
The key point of our approach is that it is valid 
for the full spectrum of invariant masses produced in the final state, and thus for
high masses final states, like the
production of a pair of $W$ bosons or the Higgs boson. 
This allows a direct comparison with experimental results already obtained at the LHC beyond the electroweak scale,
where a very good agreement has been observed between our expectations and the measurements.

Finally, we remind that these calculations are commonly used for the physics case of heavy-ion collisions.
For example, this already exists with some approximations in the {\sc Starlight} Monte-Carlo, 
mainly focused on the low invariant mass domain around the mass of the $J/\Psi$.
A complication, properly taken into account in {\sc Starlight}, arises in such collisions, due to the large value of the charges of the ions.
Then, photon-photon interactions may be accompanied by additional electromagnetic reactions,
such as photo-nuclear interactions, and the ions that come out from the collisions may be produced with some neutrons.

\section{Acknowledgements}
This work was partly supported by the Polish National Science Centre under contract No. UMO-2012/05/B/ST2/02480.


\end{document}